\documentclass[11pt,twoside]{article}
\usepackage{asp2010}
\usepackage{natbib}

\resetcounters

\begin{document}

\title{Extracting information from the data flood of new solar telescopes. Brainstorming}
\author{A. Asensio Ramos$^{1,2}$
\affil{$^1$Instituto de Astrof\'{\i}sica de Canarias, 38205, La Laguna, Tenerife, Spain}
\affil{$^2$Departamento de Astrof\'{\i}sica, Universidad de La Laguna, E-38205 La Laguna, Tenerife, Spain}}

\begin{abstract}
Extracting magnetic and thermodynamic information from spectropolarimetric observations
is a difficult and time consuming task. The amount of science-ready data that will be
generated by the new family of large solar telescopes is so large that we will be forced
to modify the present approach to inference. In this contribution,
I propose several possible ways that might be useful for extracting the thermodynamic
and magnetic properties of solar plasmas from such observations quickly.
\end{abstract}

\section{Introduction}
In the last decades, night-time telescopes have systematically increased the diameter of 
the primary mirror. Today, several facilities of the 10m class (VLT, Keck, GTC, ...) are producing top quality
science and projects for 20-40m class telescopes are already very advanced. On the contrary, the diameter of solar telescopes 
has not increased much in the last decades. Today, the most successful telescopes have primary mirrors 
in the range from 40 cm to 1 m and they exist since the 90s. The new generation of solar
telescopes has increased the size of the primary mirror to the 1.5m class (NST, GREGOR). This slow increase
in the diameter is partly motivated by the technical difficulty of heat rejection that is posed when a very large
primary mirror is used. However, the ATST (in the very early phases of construction) and EST 
(in design phase) telescopes, belonging to the 4m class, will open a completely new 
window to the investigation of the Sun.

Such large photon collectors, when combined with advanced instrumentation, will allow
us to obtain observations of the solar atmosphere with extraordinary spatial resolution, 
temporal cadence and polarimetric sensitivity. Two-dimensional spectropolarimetry on
2k$\times$2k cameras on dozens of wavelengths with a time cadence of several seconds during observing periods
of hours (thanks to the presence of multi-conjugate adaptive optics systems) will
be custom. Such amount of data (especially when several instruments work simultaneously, which is
one of the requisites of the new 4m class telescopes) will challenge the designers of 
the storage infrastructure. Even more complicated is the challenge posed to the researchers
in charge of extracting physical information from the observations. Although not all observations 
will be analyzed with inversion codes, the inversion community has to have in mind the enormous
amount of data produced by such instruments and face the challenge of developing inversion tools 
that can extract thermodynamic and magnetic properties at such rate. 
Just as an example, an instrument with a 2000$\times$2000 camera observing one or
several spectral lines during three hours at a cadence of one observation per minute
will produce 720 million line profiles for later analysis. If we consider an inversion
method that can invert one profile per second, one would need 22.8 years to invert the full dataset (and
that is only 3 hours of observations). This figure goes down to 8.3 days if an inversion is done every millisecond (or if
1000 processors are working on the problem with 1 s inversions).

\begin{figure}[!t]
\plotone[width=0.75\textwidth]{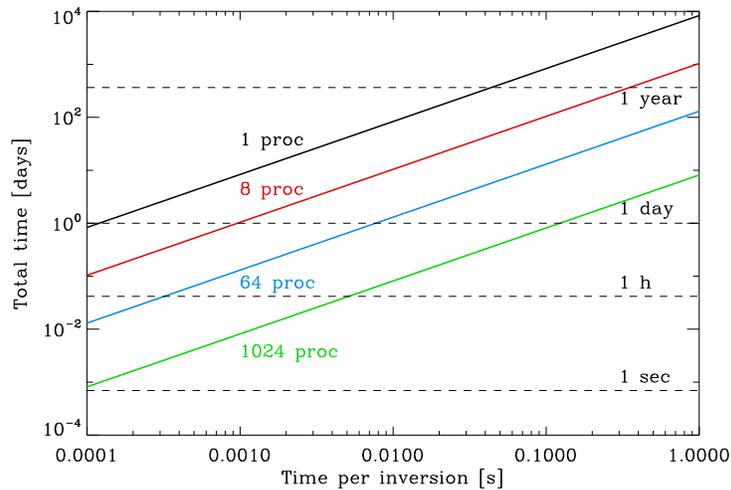}
\caption{Total computing time for
the inversion of 720 million profiles depending on the time for one single inversion. The 
solid color lines indicate the computing time for computer clusters of different sizes.}
\label{fig:computing_time}
\end{figure}

A summary of the previous computations is shown in Figure \ref{fig:computing_time}. The color
lines in the figure represent the total amount of time (in days) needed for inverting 720 million 
profiles depending on the computing time per inversion (in seconds). One can see that, if
only one processor is used, the total computation time goes down to one day if the computation time per inversion 
is as low as 0.1 ms. If one wants to carry out a systematic inversion of all observations
(assuming an average of 3 hours of good observing time per day) on a day-to-day basis, inversions
need to be carried out in less than 0.1 ms. Equivalently, if a dedicated cluster of 1000 processors
is used for this purpose, the time per inversion can be safely increased to 0.1 s without too
much impact.

Such reduced computing times per inversion can be reached with fast computers for simple 
Milne-Eddington atmospheres using optimized codes like VFISV \citep{borrero07,borrero_vfisv10}
tailored to the inversion of a very specific dataset. This code is presently used for the inversion 
of filterpolarimetric data of the Helioseismic and Magnetic 
Imager (HMI; onboard the Solar Dynamics Observatory). One can expect that the computing power
increases following Moore's law and that computers at the time when ATST and EST
start observing can be a factor 8-32 faster (computers duplicate the number of
transistors every 2-3 years) than present day computers. Another possibility is
to consider the application of graphical processing units (GPU) to the inversion
problem. The advantages of GPUs rely on the reduced price and the enormous parallelization
capabilities for data-parallel computations. It remains to be investigated to what extent GPUs
can accelerate the inversion process. An additional option is to use hardware solutions like
FPGAs (field programmable gate arrays) that can be configured for the inversion process (of
relatively simple models because of the complexity of programming such devices). This option
is considered for the onboard inversion of data for the Polarimetric and Helioseismic Imager (PHI)
onboard Solar Orbiter.

The main difficulty with all these approaches lies in the fact that the inversion of chromospheric
lines (which are in the mandatory list of spectral lines for both ATST and EST) is
not as simple as the inversion of photospheric lines. Inversion of photospheric lines
can be easily done in local thermodynamical equilibrium (LTE), which speeds up the
computations. On the contrary, chromospheric lines tend to have large non-LTE
corrections and the full radiative transfer problem has to be solved in each step of
the iterative inversion process. This drastically increases the computation time and
it is very complicated to do inversions in less than a few seconds. This is the
case of NICOLE (Socas-Navarro, de la Cruz Rodr\'{\i}guez,
Asensio Ramos, Trujillo Bueno \& Ruiz Cobo, in preparation), to my knowledge, the only
inversion code that solves the full non-LTE problem (neglecting the presence of
atomic polarization). Additionally, these
lines are formed in regions where scattering effects are important and, consequently,
the polarization signals have strong contributions (or are even dominated) by atomic
polarization induced by anisotropic pumping. The inversion of Stokes profiles
dominated by atomic polarization is even more computationally intensive and
the only existing codes that can cope with such a problem are Hazel \citep{asensio_trujillo_hazel08} and
Helix+ \citep[an updated version of the code used by][]{Lagg04,lagg07}, both based
on a Levenberg-Marquardt (LM) algorithm, and the code of \citep{lopezariste_casini05} based on
a look-up table. Hazel has been recently parallelized to cope with large-scale inversions,
scaling roughly linearly with the number of processors.

The previous considerations force us to consider how to approach the inversion of photospheric and
chromospheric lines with large telescopes. I will consider in this contribution several
options that we would need to investigate in depth (and other options that can be
derived from them) and that can help us face the problem with success.

\section{Classification}
The conclusion from the previous paragraphs is that the number of profiles that
we need to invert is extraordinarily large. We have to change the paradigm and
admit that we cannot look at all data (we lack the manpower for
that even if the computing power is available). A possible option to overcome the
difficulties is to apply algorithms that reduce the number of profiles to something
we can deal with. We can develop automatic
algorithms that can classify the observed profiles as intrinsically interesting
or uninteresting. These algorithms have to work online in the telescope and either
raise flags to classify them for later consideration or just throw away all 
uninteresting information (which is probably not welcomed by the solar
community, although it is the standard procedure in other fields like in
experimental particle physics). These online algorithms should ideally be based
on supervised classification that need to be trained in advance. Therefore, we first have to solve
the problem of defining what is interesting and what is not. Technically, 
we can choose among artificial neural networks (ANN), support vector machines (SVM) or Gaussian
processes (GP). All of them have demonstrated robust classification capabilities once
trained appropriately.

\begin{figure}[!t]
\plottwo{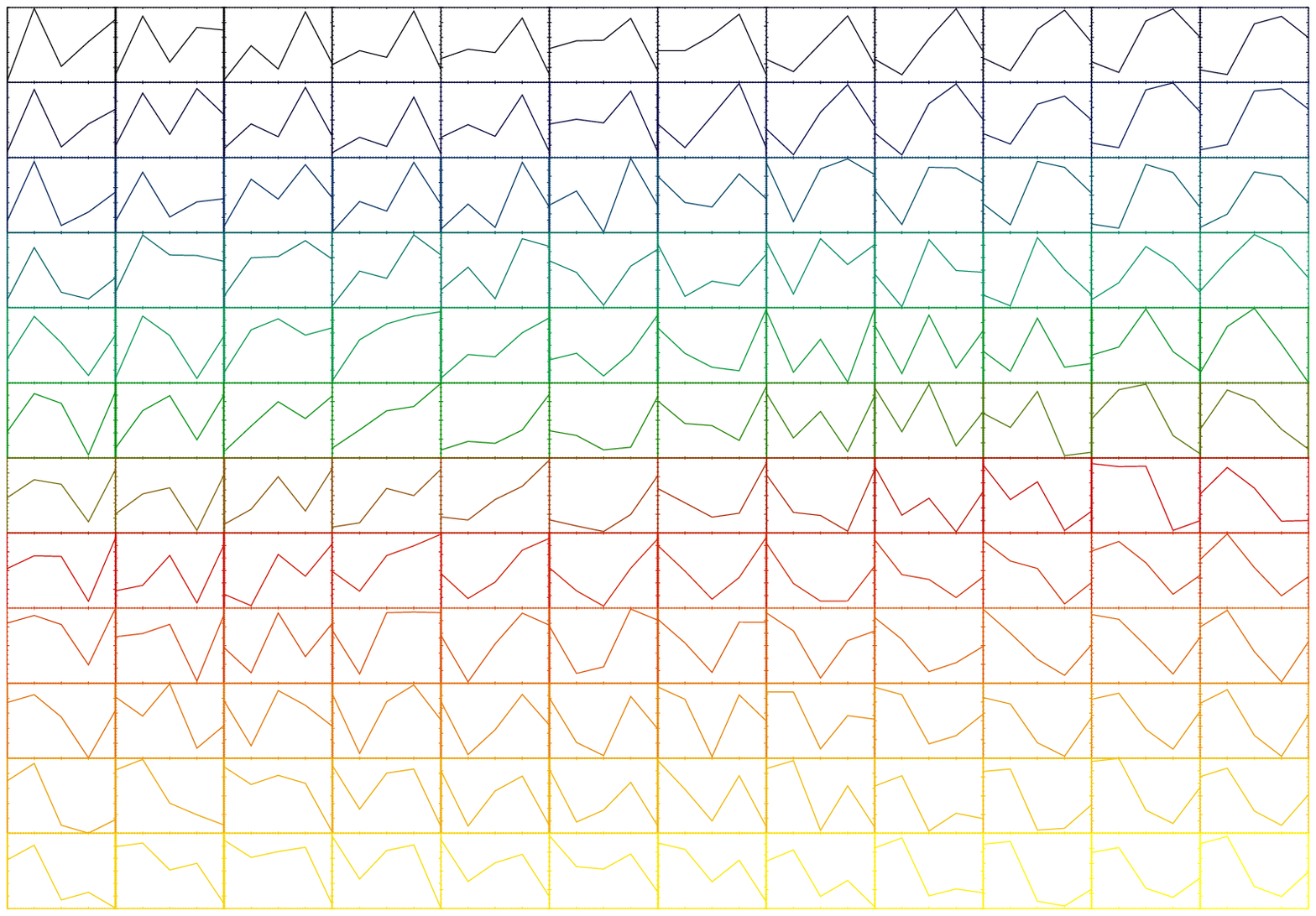}{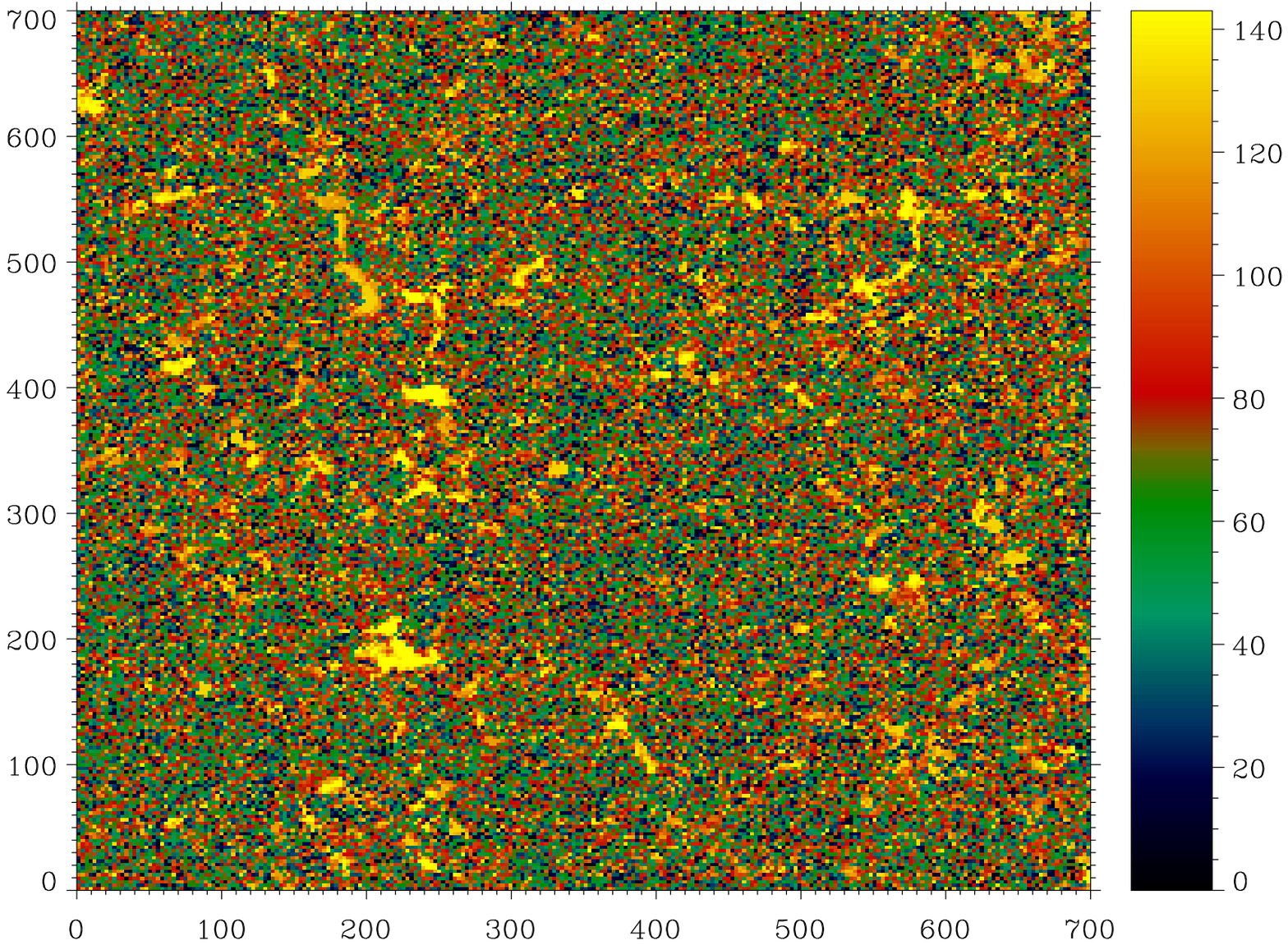}
\caption{The left panel shows a 12$\times$12 self-organizing map applied to the Stokes $V$ profile
of a snapshot of 5-point IMaX data. The SOM has classified similar profiles in nearby regions. By
transforming the five-dimensional space of IMaX profiles to a two-dimensional space, we can
easily identify anomalous profiles. The right panel displays in colors the class (from the 144 possible classes)
to which every profile in the map belongs.}
\label{fig:som_imax}
\end{figure}

A less aggressive option is to consider unsupervised classification and dimensionality
reduction methods once all the data has been stored. A line profile sampled
at $n$ wavelength points can be represented as a point in the $n$-dimensional space $\mathbb{R}^n$.
Although this space can have very large dimensionality, physics and the measuring
instrument incorporates correlation among the different wavelength points. Consequently,
the effective dimensionality of the measured Stokes profiles is decreased because they
``live'' in a low-dimensionality manifold of the $n$-dimensional space. Therefore, dimensionality 
reduction methods try to capture this behavior and present the user with a simplified
version of the observations. The self-organizing map (SOM) belongs to this category of
dimensionality reduction methods. It is an ANN where each neuron is characterized by an $n$-dimensional
vector that represents a Stokes profile. It is trained using
unsupervised learning techniques to produce a low-dimensional representation of the training sample. This
representation tries to preserve the topological properties of the input space (training samples), so 
that nearby samples are mapped into nearby neurons 
\citep[][]{kohonen_SOM01}. One of its most important applications is for visualization of high-dimensional data.
This method has been used by \cite{asensio_mn07} to classify profiles of the Mn \textsc{i} line at 15262.702 \AA.
I refer to that paper for more details. After convergence of the map, the profile associated to each neuron tends
to be associated with \emph{patterns} in the input data, with similar patterns being located in nearby
neurons. Consequently, once trained with some data, it is possible to use the map to classify any additional
input vector, whether it was in the training set or not. The euclidean distance is calculated between the input 
vector and all the profiles associated to the neuron. The neuron closest to the input vector will give the class.
An example of this procedure is shown in Fig. \ref{fig:som_imax}, where we have trained a SOM with
Stokes $V$ profiles obtained from IMaX observations \citep{imax11} onboard the Sunrise balloon \citep{sunrise10}. The
left panel displays the classes in a 12$\times$12 array, while the right panel shows the class to which
each pixel is associated. Anomalous profiles can be identified in the SOM and their physical locations 
easily discovered.

\section{Look-up databases}
In addition to the computational problem all inversion codes based on
the LM algorithm suffer from, they also have problems
defining realistic error bars, specially when ambiguities are present. Realistic
error bars can be obtained using a fully Bayesian approach \citep{asensio_martinez_rubino07},
with the inconvenience of increasing the computational problem because a Montecarlo approach
(thus requiring many evaluations of the forward problem) is used to sample the posterior distribution.

One approach that has been demonstrated to partially solve both problems is the
one of building large databases. The database contains many Stokes profiles evaluated
at all physically relevant combination of the parameters of the model under
consideration. A direct comparison of the observed profile with all the profiles present
in the database gives us an estimation of the model parameters. Additionally, error bars (although
not fully Bayesian) can be estimated taking into account all profiles inside a ball around the best 
model. The database approach
works well although the inherent flexibility of LM codes has to be sacrificed because the
database has to be built in advance. The database approach has been pursued by \cite{arturo_casini02}, \cite{casini05},
\cite{lopezariste_casini05} and \cite{casini09} with great success. Its application to
the very complex problem of scattering polarization and the Hanle effect is specially
relevant because the database needs to be computed only once and then can be applied without
much computational burden to the observations.

Building the database is not an easy task because it might suffer from the curse of dimensionality. This is
related to the fact that, when the dimensionality of a space increases, the volume of the space increases 
so fast that any available sampling becomes sparse. In other words, the size of a database should increase
exponentially fast with the number of wavelength points of the sampled Stokes profiles. Fortunately, this
is partially solved because the assumed parametric models that generate the Stokes profiles efficiently
reduce the dimensionality of the problem (inducing that the Stokes profiles lay in a low-dimension manifold of the 
full space). Casini, L\'opez Ariste and co-workers have developed Montecarlo algorithms to efficiently
build the database. Additionally, instead of storing the full Stokes parameters in the database,
they project them to a small subset of principal components, gaining compression and reducing the
dimensionality of the problem.

When the mapping between the input model parameters and the output Stokes parameters is complex
(like when scattering polarization and the Hanle effect play a role) and the noise level of
the observations is small, the size of databases has to be increased dramatically. Above a
certain size of the database, the search algorithm turns out to be slower than an inversion
code based on the LM algorithm like Hazel. For a database with $n$ Stokes profiles, the search 
scales as $O(n)$. There is not much room for improving the search because only a few methods
seem to be significantly better than a brute-force computation of all distances. 
However, by relaxing the problem and computing nearest neighbors approximately, it is possible to 
achieve significantly faster running times (of the order of 1-2 orders of magnitude) often with a relatively small actual errors.

\begin{figure}[!t]
\plottwo{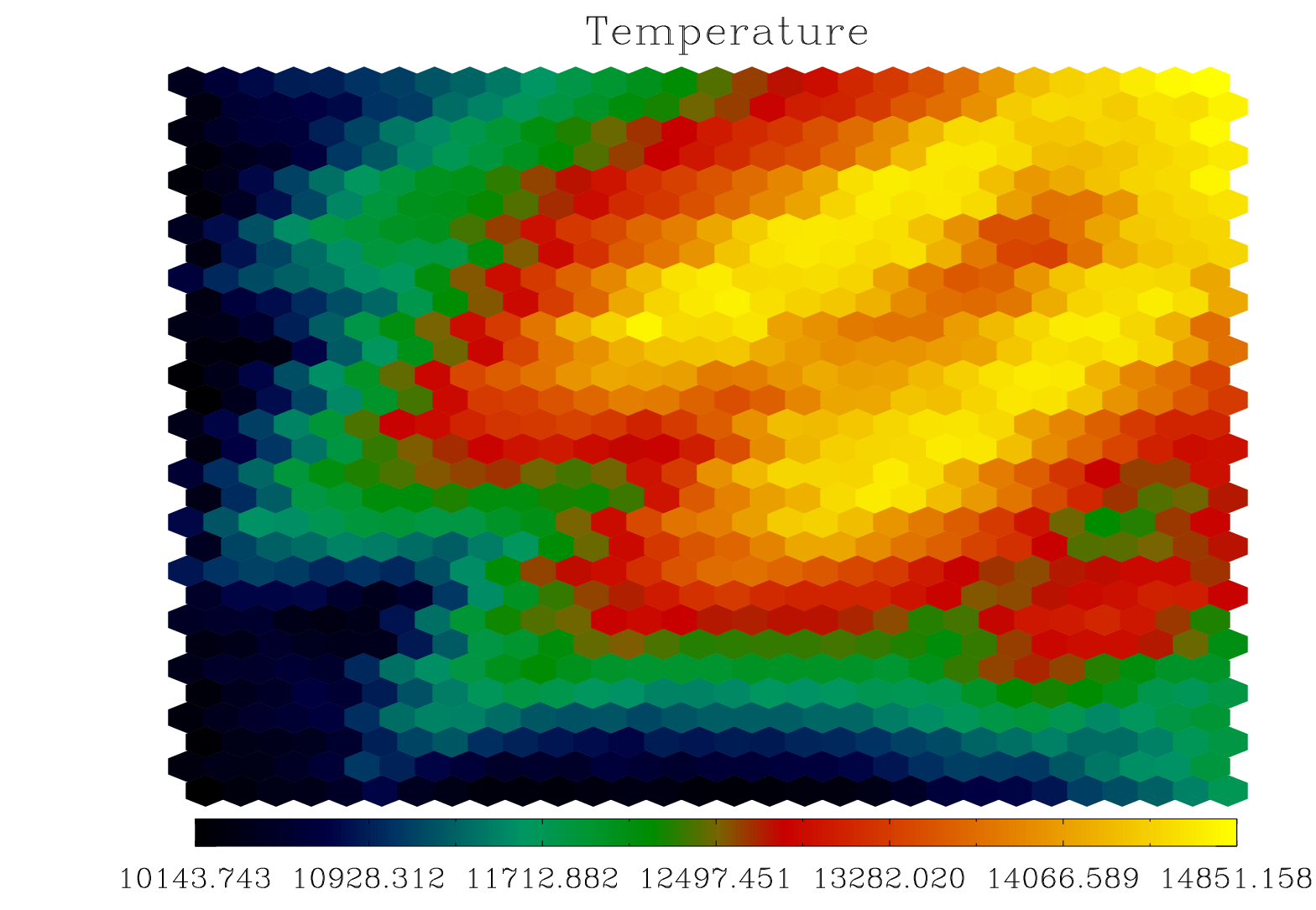}{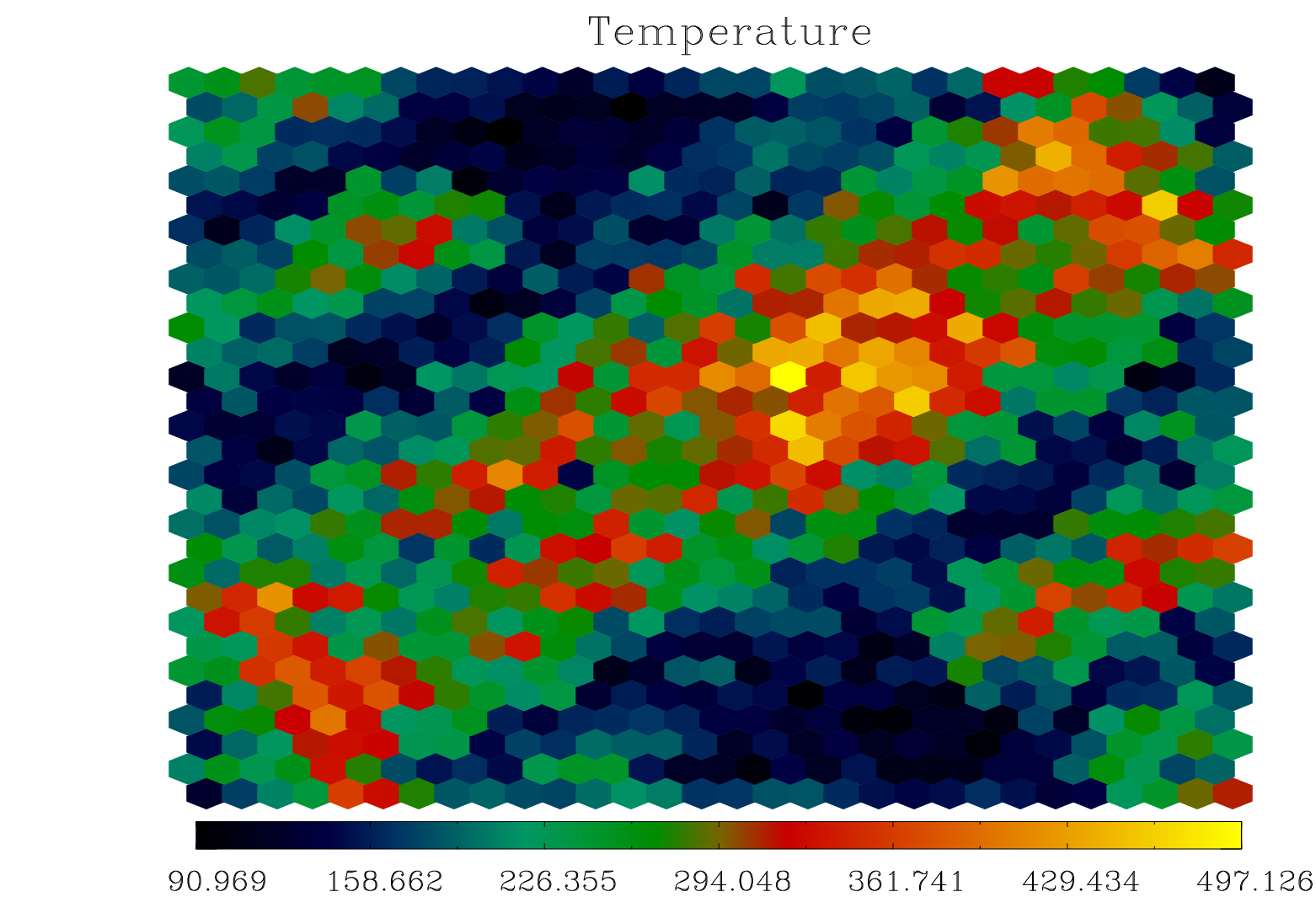}
\plottwo{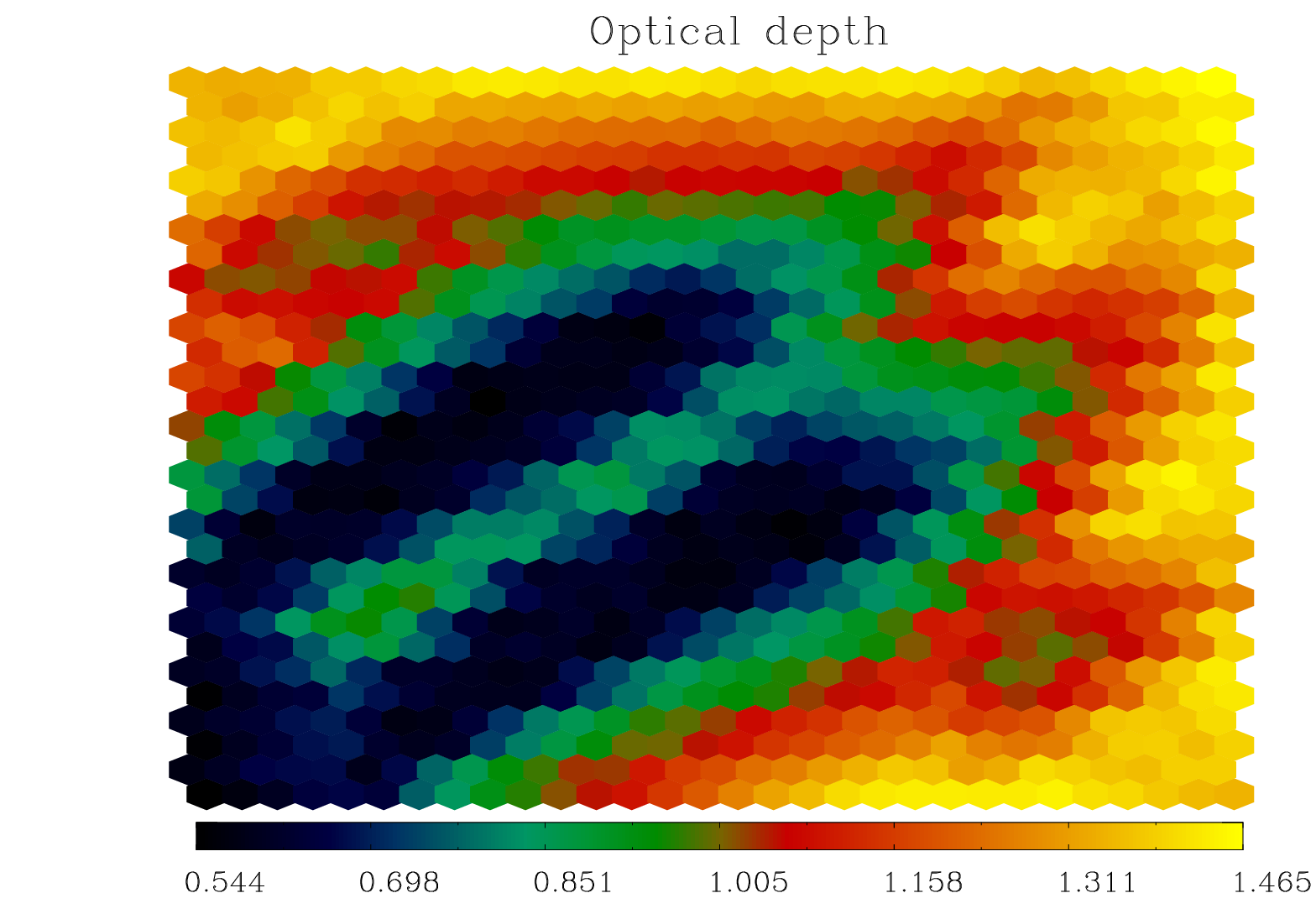}{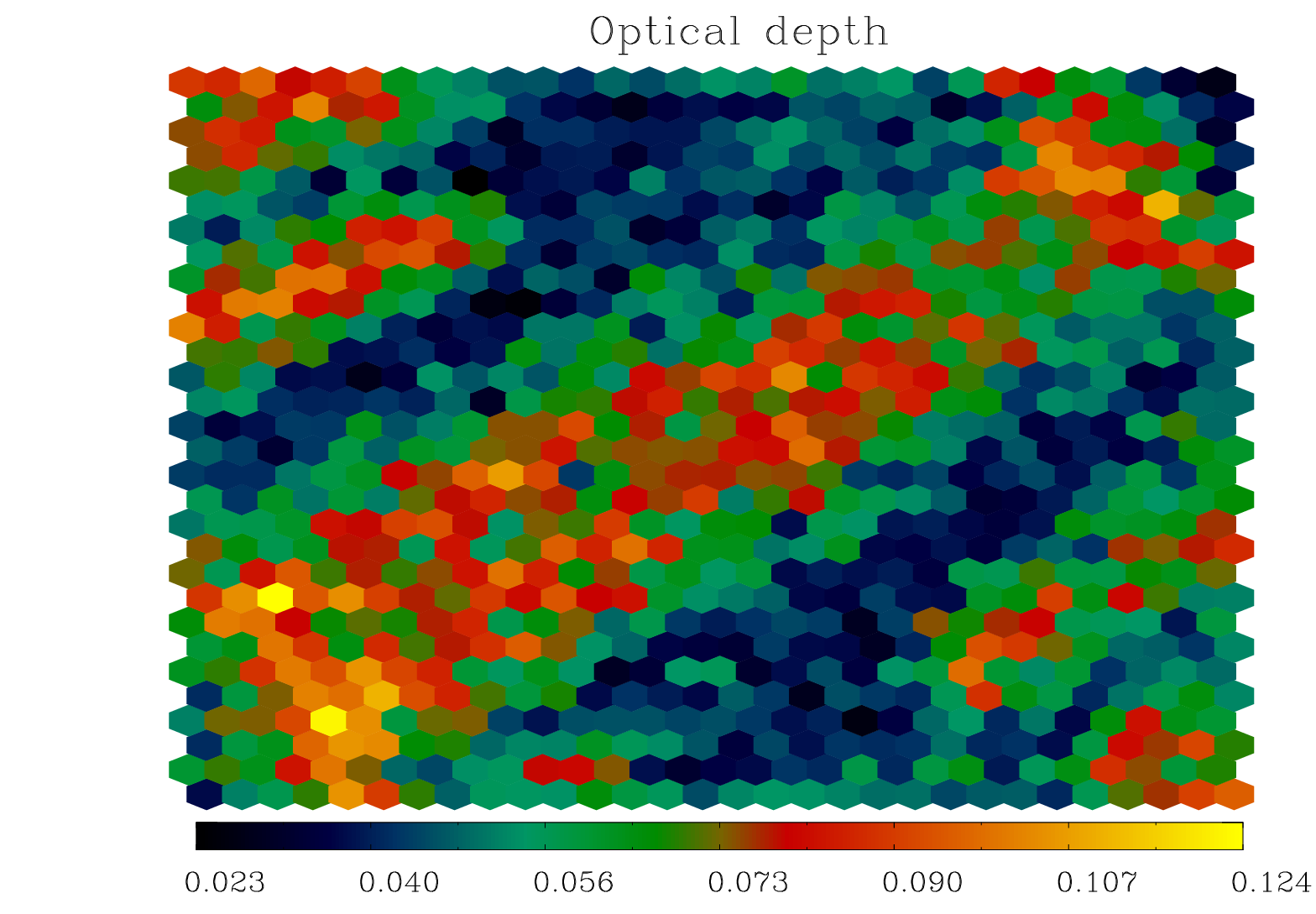}
\plottwo{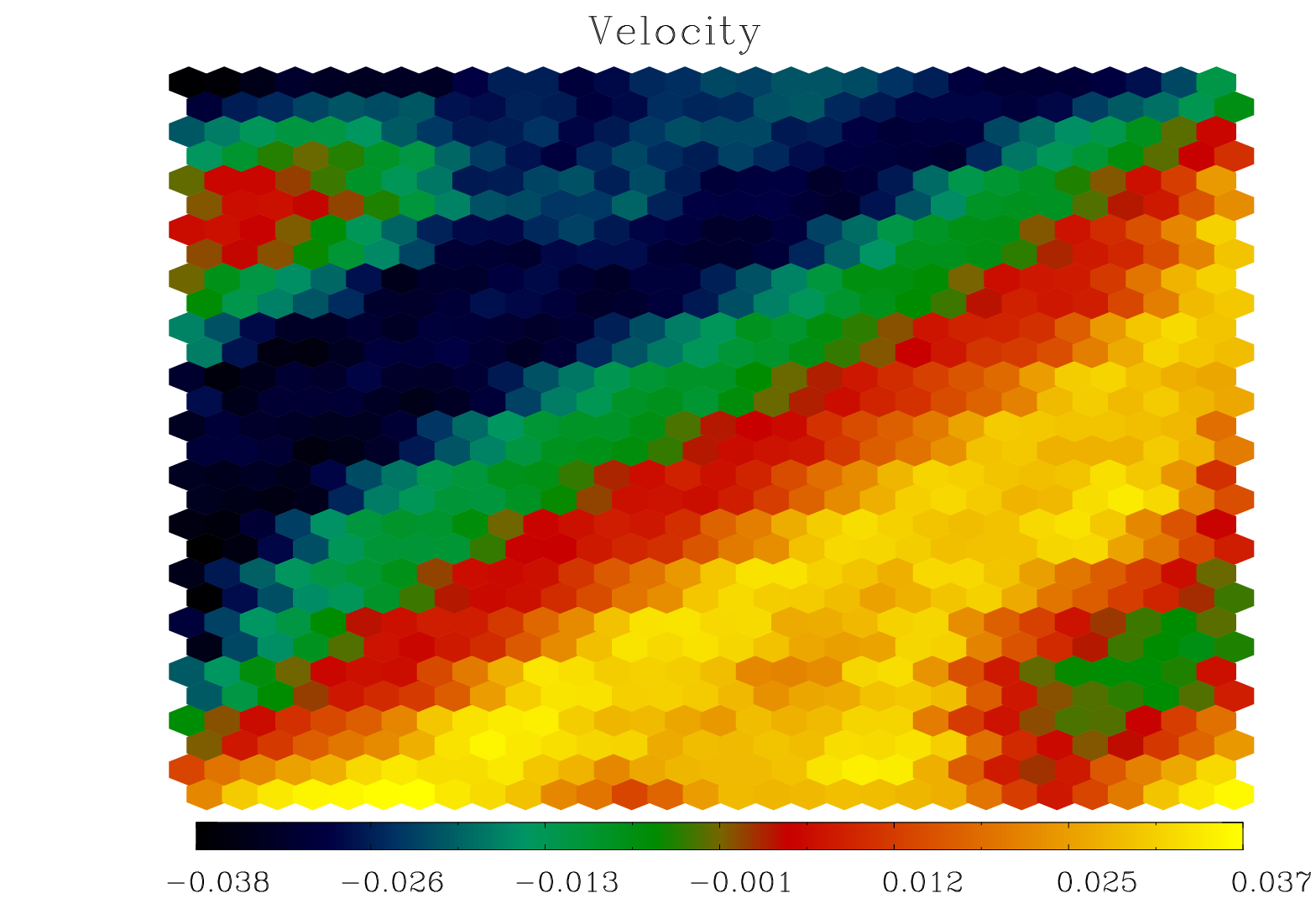}{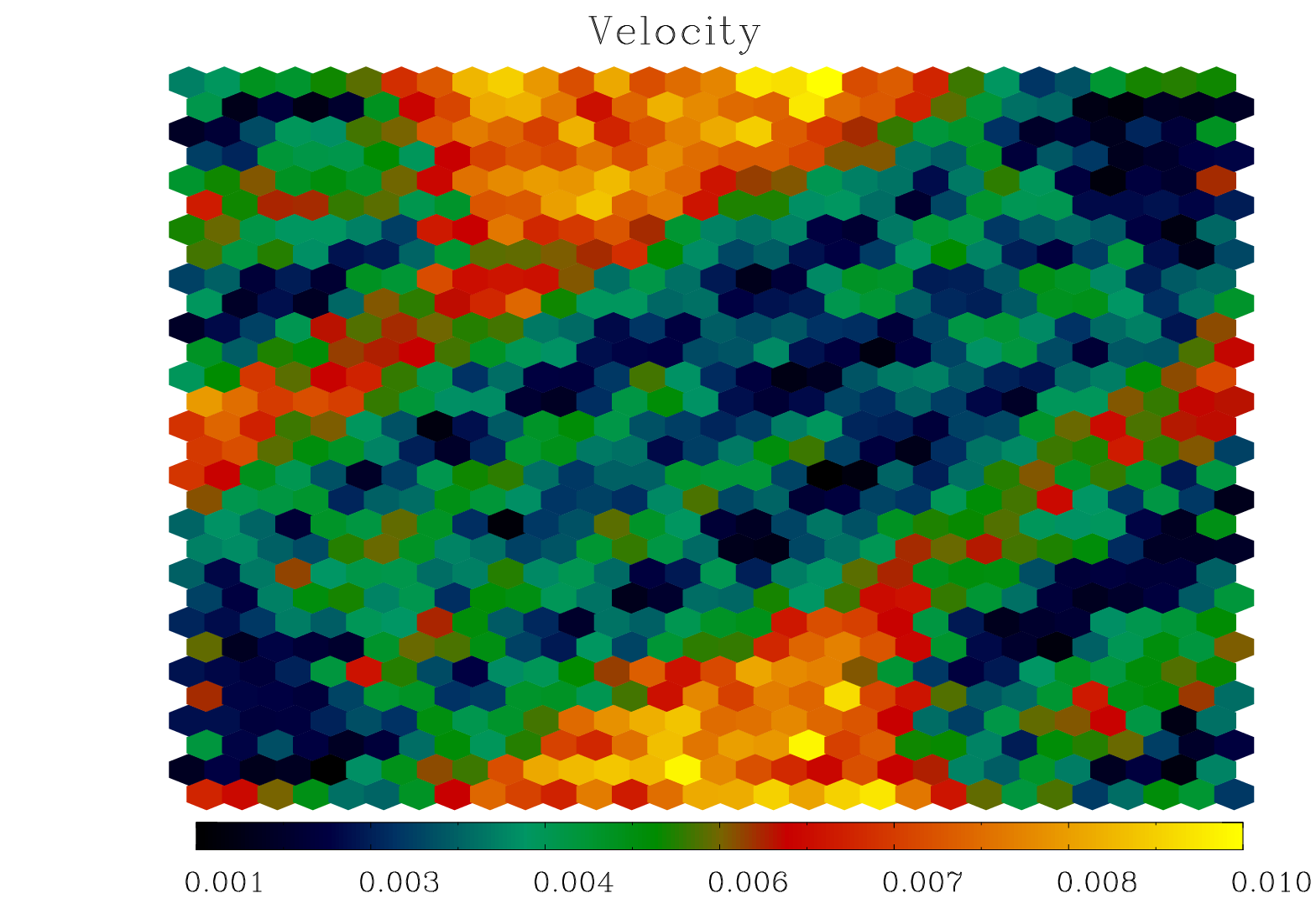}
\caption{The left columns show the temperature (upper panel, in K), optical depth (middle panel) and velocity (lower panel, in km s$^{-1}$) associated
to each neuron of the SOM trained on a set of Stokes $I$ profiles of the 10830 \AA\ multiplet. It is computed as
the mean of all profiles in the training database that are associated with each neuron. The right panels display
the associated standard deviation. This SOM can be used to quickly estimate the physical parameters from
the observables.}
\label{fig:som_helium}
\end{figure}

An option for very large databases is to reduce its size using a SOM and use it
for inversion purposes to give a rough estimation of the model parameters.
To show this, I have trained a 30$\times$30 SOM using the Stokes $I$ profiles for the 10830 \AA\ multiplet kindly provided
by R. Casini from the database used by \cite{casini09}. We extract the temperature, optical depth and
velocity of each one of the profiles associated to a given neuron, compute its average and the standard
deviation. The results are shown in Fig. \ref{fig:som_helium}. This demonstrates that, presenting a new
observation to the SOM and picking up the neuron giving the smallest euclidean distance, gives an estimation
of the temperature, optical depth and velocity with a small error associated. One of the advantages of
this fast inversion is that one only needs to compare the input Stokes profiles with 900 profiles, instead
of the full database. If only approximate values of the physical parameters are desired, this might
give a large improvement in terms of computation work.

\section{Emulators}
We have stated that standard inversion tools are quite slow when the forward problem is complex.
A possible solution to this issue lies on the application of emulators. We call an emulator any
machine learning method that ``learns'' the mapping between the thermodynamic and
magnetic parameters (expressed in vector form as ${\mbox{\boldmath$\theta$}}$) and the emergent 
Stokes profiles, $[I(\lambda),Q(\lambda),U(\lambda),V(\lambda)]$. The main advantage of this approach is
that, if the selected machine learning method, once trained, is faster than solving the full 
problem, we can apply a standard LM algorithm to carry out the inversion. Additionally, 
increasing the speed of the computation of the forward model opens up the option
of carrying out Bayesian inversion through the use of efficient Markov Chain Montecarlo
methods. ANNs, GPs and SVMs are standard machine learning methods that can be used
to learn the mapping. In order to
simplify the problem, the Stokes profiles can be decomposed as a linear combination of the principal components,
so that the mapping to be learnt is between ${\mbox{\boldmath$\theta$}}$ and the projection of every
Stokes profile along the principal components.

The main difficulty of this approach resides on the precision that needs to be imposed to the
mapping. If the noise in the observation is characterized by a standard deviation $\sigma$, the machine
learning method needs to synthesize profiles that are, for a set of thermodynamic and
magnetic parameters, closer than $\sigma$ to the correct synthetic profiles in order not to introduce
artificial biases. This is a very complicated task and more work needs to be done. As
an example, \cite{bayesclumpy09} uses an ANN to synthesize spectral
energy distributions of the clumpy dusty tori models of \cite{Nenkova08a}. The restrictions
put to these ANNs are very relaxed given the large observational errors. The synthesis of Stokes
profiles is much more restrictive in terms of precision.

One possible improvement in the quality of emulators resides in modifying the decomposition
of the Stokes profiles. The projection along the principal components is a linear
transformation in the space of Stokes profiles. It can be understood as a rotation
in an euclidean space and a projection along a low-dimensional hyperplane. As a consequence, 
it is not able to capture the shape of the manifold where the Stokes profiles live. This
poses more difficulties to the machine learning method that needs to capture the
non-linear behavior of the manifold. Fortunately, more elaborate non-linear dimensionality
reduction techniques like diffusion maps \citep{coifman_lafon06,lafon_lee06}, locally linear embedding 
\citep{lle00}, isomap \citep{isomap00}, kernel-PCA \citep{kpca98} and autoassociative 
artificial neural networks \citep{socas_navarro05} exist. In principle,
one could think that allowing the dimensionality reduction to capture part of the
non-linearity of the manifold will help any machine learning method to learn
the mapping between the input parameters and the Stokes profiles. This possibility needs to be
investigated in more detail. The same strategy could also be applied for improving 
machine learning methods that learn directly the mapping between the Stokes profiles
and the physical parameters.

\section{Model selection}
As stated above, inferring magnetic and thermodynamic information from spectropolarimetric observations
relies on the assumption of a parameterized model atmosphere whose parameters are tuned by comparison with observations.
Often, the choice of the underlying atmospheric model is based on subjective reasons.
In other cases, complex models are chosen based on objective reasons (for instance, the
necessity to explain asymmetries in the Stokes profiles) but it is not clear what degree of complexity
is needed. The lack of an objective way of comparing models has, sometimes, led to opposing
views of the solar magnetism because the inferred physical scenarios
are essentially different. This can be solved using Bayesian model selection tools, allowing
us to determine which is the model best suited 
for explaining the Stokes profiles observed in a pixel \citep[e.g.,][for a general
description and more details]{trotta08}.
Let's assume we have $N_\mathrm{mod}$ models $\{ \mathcal{M}_i,i=1\ldots N_\mathrm{mod}\}$ competing 
to explain the same set of observations formally represented by $D$. 
Here, $D$ will be represented by a formal vector $\mathbf{d}=[I(\lambda_1), I(\lambda_2), ..., Q(\lambda_1), ..., U(\lambda_1), ..., V(
\lambda_1), ...]$ whose elements are the values of the Stokes parameters $I$, $Q$, $U$, and/or $V$ at certain wavelengths $\lambda_1, 
\lambda_2, ...$.
By a \emph{model} we mean an algorithm that depends on a set of $N_j^{(i)}$ parameters ${\mbox{\boldmath$\theta$}}_i=(\theta_{i; 1}, \theta_{i; 2}, 
..., \theta_{i; N_{j}^{(i)}})$ (often, the temperature at one or several points in a model atmosphere; the magnetic field strength, in
clination, and azimuth; the density, etc), whose output is a prediction $\mathbf{y}({\mbox{\boldmath$\theta$}}_i)$ of the data. 
The Bayes theorem \citep{jaynes03,mackay03,gregory05} states that the posterior probability of each model at the light of the obs
erved data is
\begin{equation}
p(\mathcal{M}_i|D) = \frac{p(D|\mathcal{M}_i) p(\mathcal{M}_i)}{p(D)},
\label{eq:prob_model}
\end{equation}
where $p(\mathcal{M}_i)$ is our prior belief in each model (which we will assume to be the 
same for all the models considered here; see below), while $p(D)$ is just a normalization constant:
\begin{equation}
p(D) = \sum_{i=1}^{N_\mathrm{mod}} p(D|\mathcal{M}_i) p(\mathcal{M}_i).
\end{equation}
Finally, $p(D|{\cal M}_i)$ is the evidence or marginal likelihood, which is the key 
ingredient of our model comparison, and is given by the following integral \citep[e.g.,][]{trotta08,asensio_spw6_11}:
\begin{equation}
p(D|\mathcal{M}_i) = \int \mathrm{d}{\mbox{\boldmath$\theta$}}_i p({\mbox{\boldmath$\theta$}}_i|\mathcal{M}_i) p(D|{\mbox{\boldmath$\theta$}}_i,\mathcal{M}_i).
\label{eq:evidence}
\end{equation}
The quantity $p({\mbox{\boldmath$\theta$}}_i|\mathcal{M}_i)$ is the prior distribution for the model parameters.
The quantity $p(D|{\mbox{\boldmath$\theta$}}_i,\mathcal{M}_i)$ in Eq. (\ref{eq:evidence}) is the likelihood, which is computed from the observed data. 
Assuming that the observations are corrupted with uncorrelated Gaussian random noise, then
\begin{equation}
p(D|{\mbox{\boldmath$\theta$}}_i,\mathcal{M}_i) = \prod_{j=1}^M \left(2\pi \sigma_j^2\right)^{-1/2} \exp \left[ -\frac{ \left(y_j({\mbox{\boldmath$\theta$}}_i)-d_j \right)^2}{2\sigma_j^2} \right],
\end{equation}
\citep[for mode details, see][]{asensio_martinez_rubino07,asensio_hinode09}.
It is important to note that, if a model parameter is completely unconstrained by the observed data
(so the ensuing likelihood does not depend on this parameter), the evidence does not penalize it
because it factorizes from the integral.

\begin{figure}[!t]
\plottwo{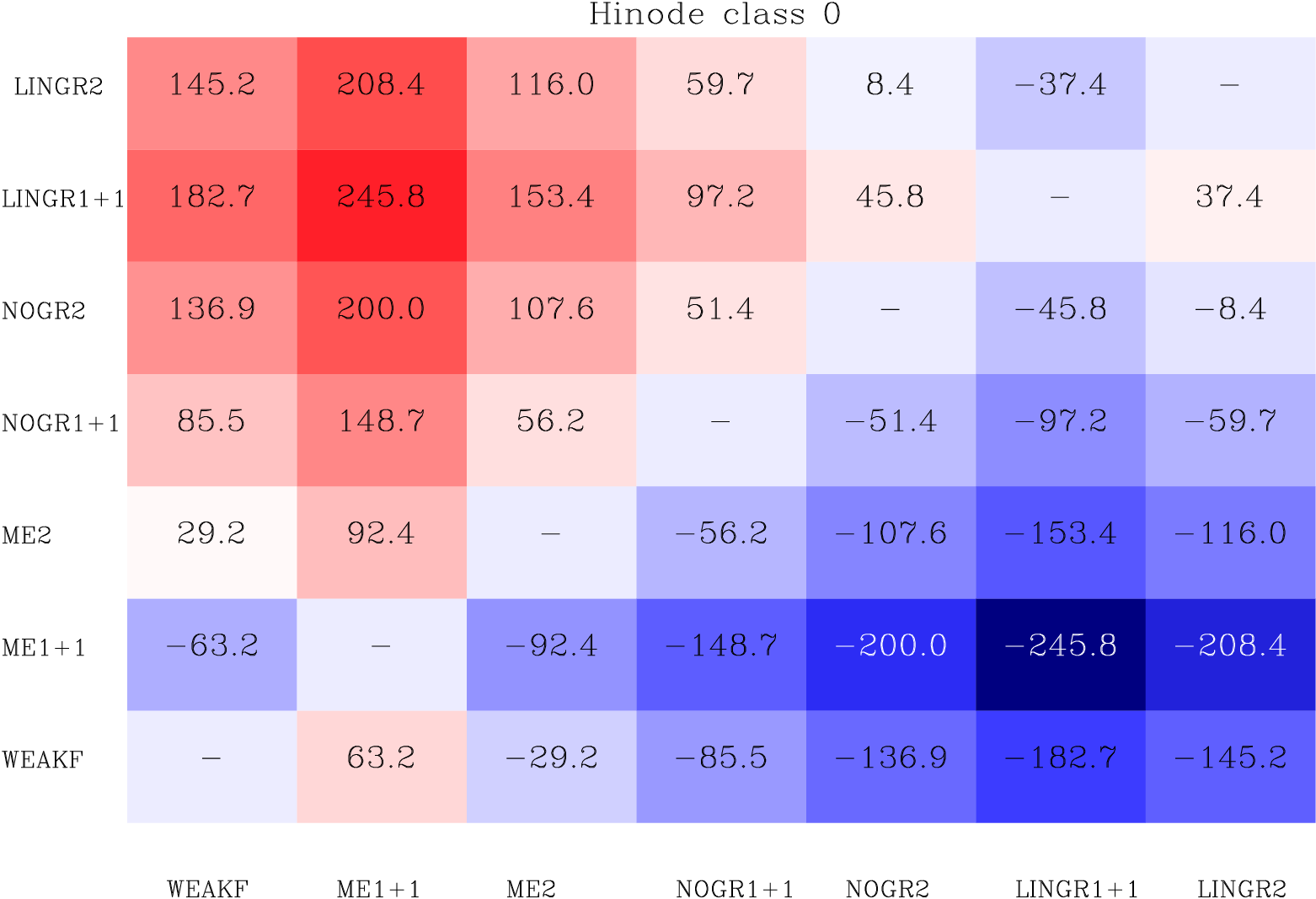}{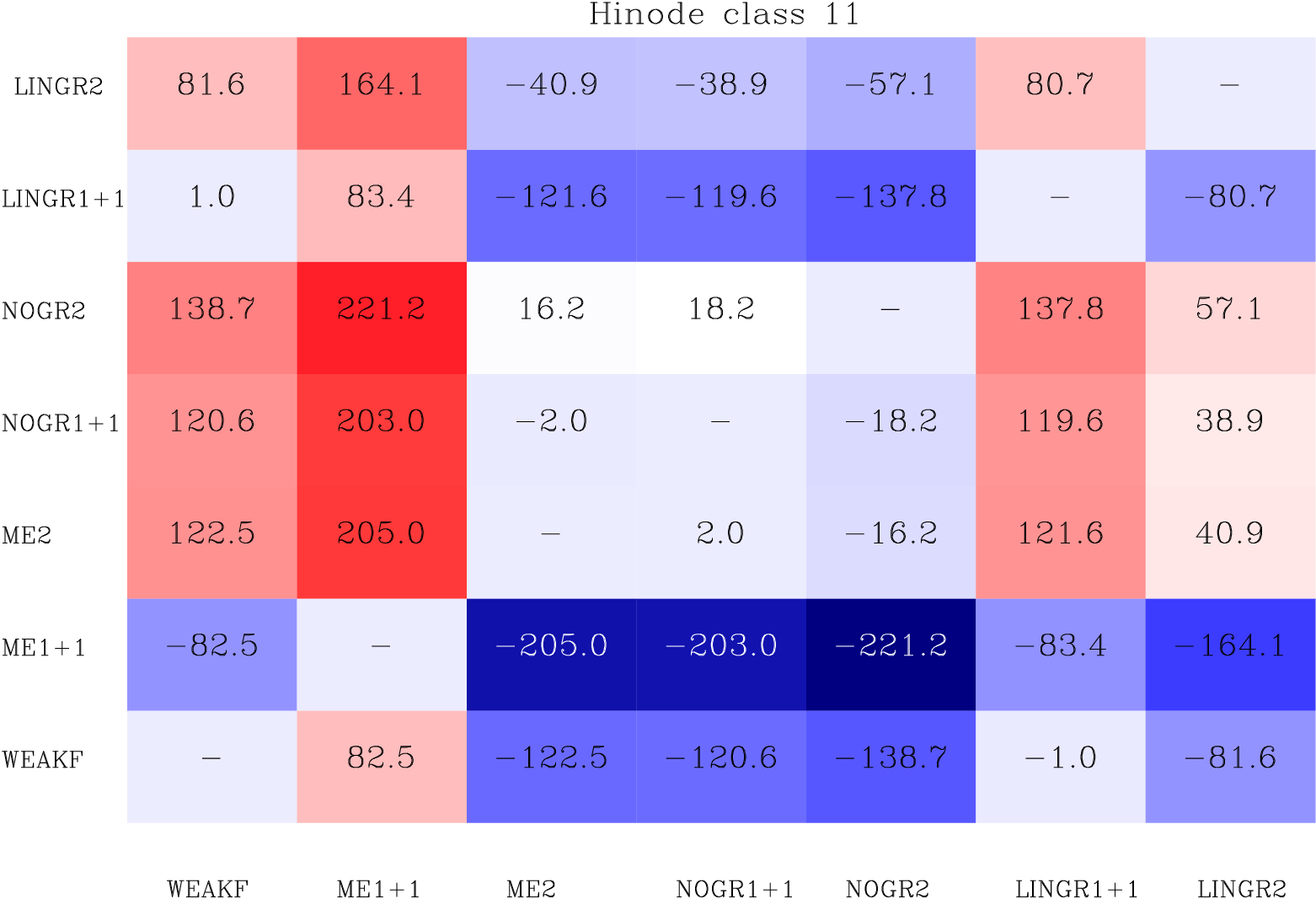}
\plottwo{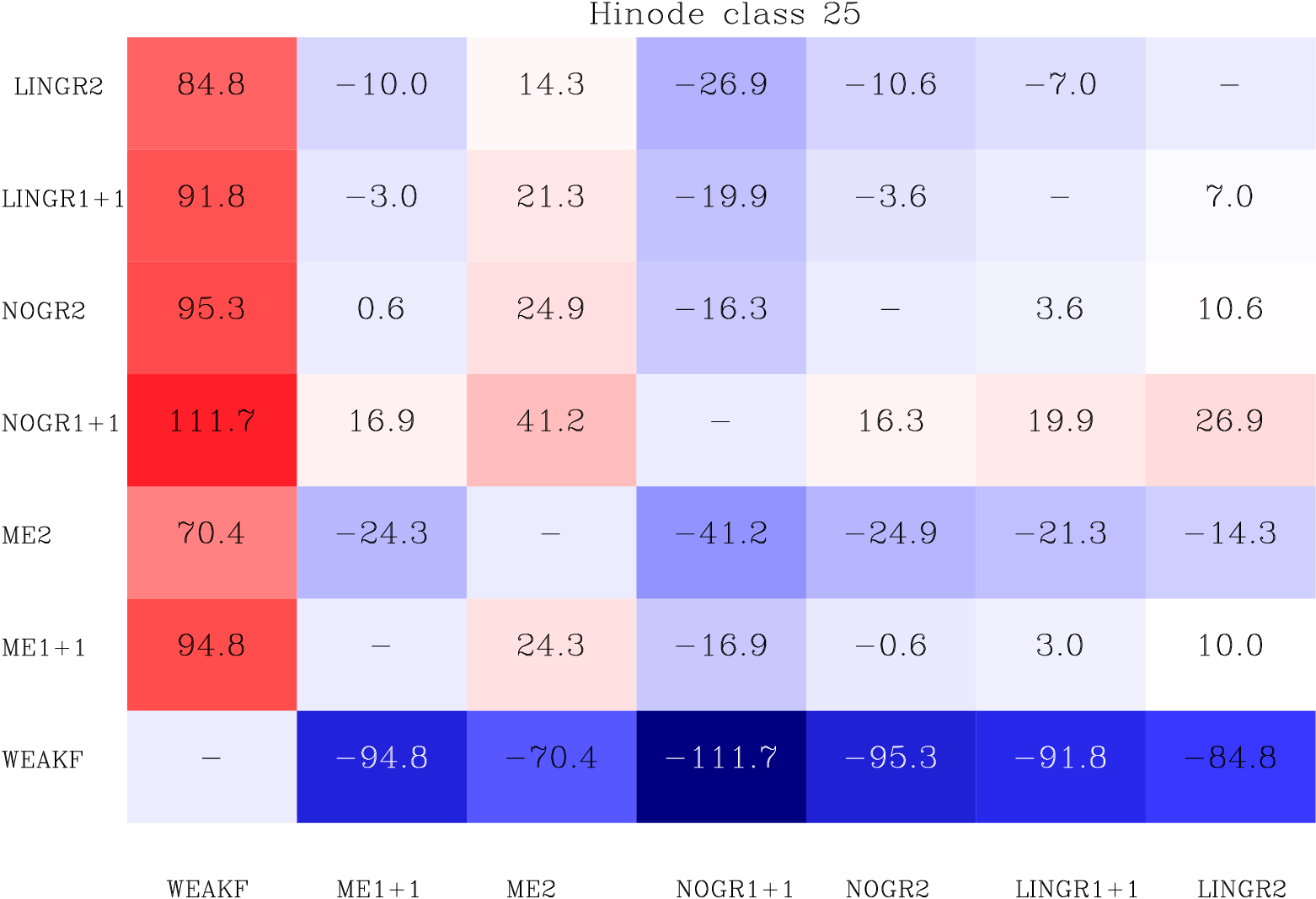}{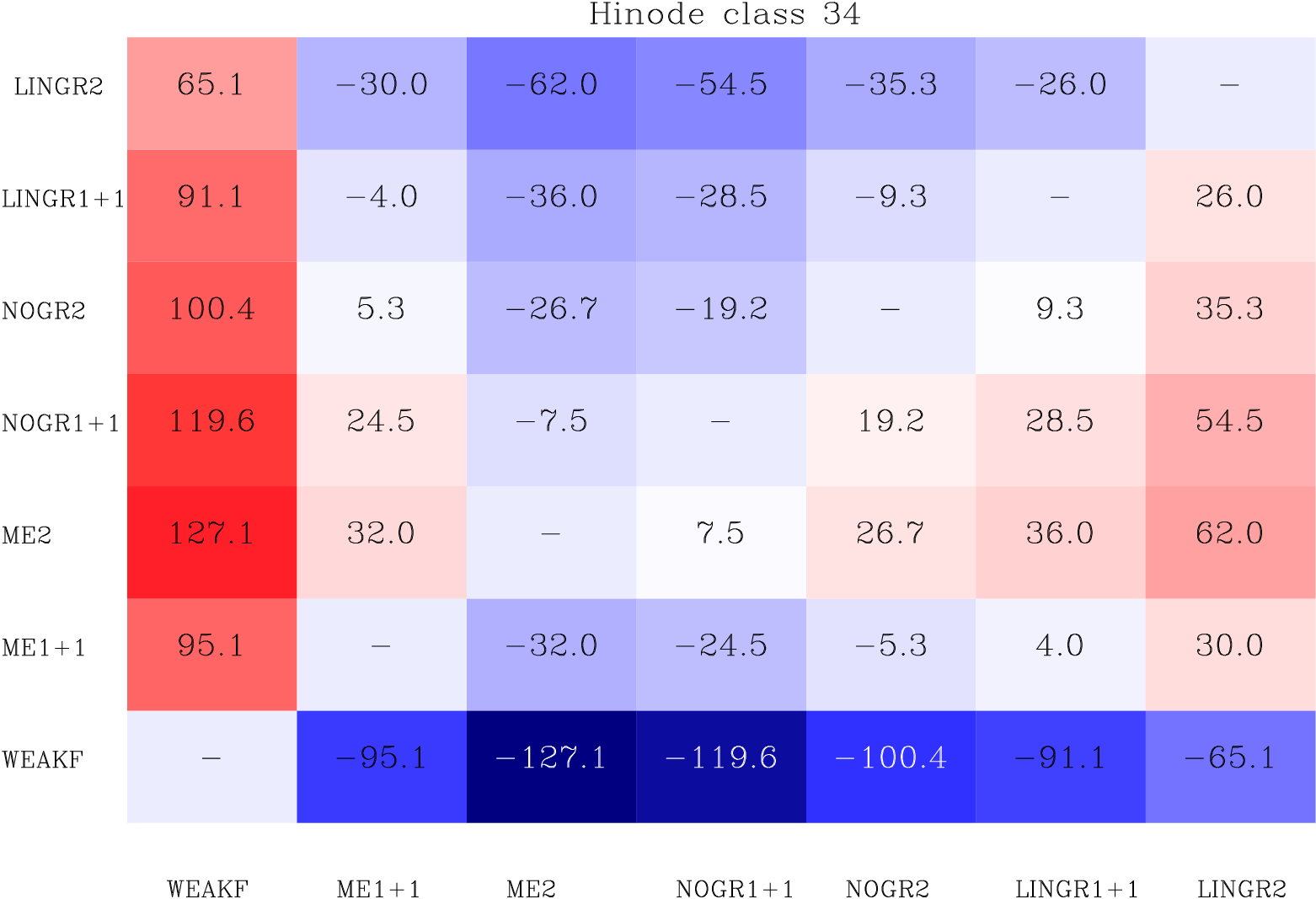}
\caption{Logarithmic evidence ratio from each model with respect to the every other model. We show
the results for four different profiles belonging to different classes as defined by \cite{viticchie_1_11}. Models can be
compared using these tables if we assume the same a-priori probability for all of them. Each square reports
the log evidence ratio between a given model in the vertical axis versus a certain model in the horizontal
axis. Red and yellow colors indicate when the model in the vertical axis is more probable and
blue when the opposite happens. Note that these tables are symmetric with respect to the diagonal.}
\label{fig:leagues}
\end{figure}

Given two models, $\mathcal{M}_0$ and $\mathcal{M}_1$ that are proposed to explain
an observation, the ratio of posteriors
\begin{equation}
\frac{p(\mathcal{M}_0|D)}{p(\mathcal{M}_1|D)} = \frac{p(\mathcal{M}_0)}{p(\mathcal{M}_1)} \frac{p(D|\mathcal{M}_0)}{p(D|\mathcal{M}_1)},
\end{equation}
is used to compute how more probable one model is with respect to the other \citep{jeffreys61}. The
ratio of evidences is known as the Bayes factor, $B_{01}$, and it is trivially given by:
\begin{equation}
B_{01} = \frac{p(D|\mathcal{M}_0)}{p(D|\mathcal{M}_1)}.
\end{equation}
If both models are assumed to have the same a-priori probability (which is what we have
assumed in all subsequent computations), the ratio of posteriors
is just the Bayes factor. Large values of $B_{01}$ indicate a preference for model $\mathcal{M}_0$
while small values indicate a preference for model $\mathcal{M}_1$. The modified Jeffreys scale 
can be used to translate values of the Bayes factor into strengths of belief \citep{jeffreys61,kass_raftery95,gordon_trotta07}.

It is illustrative to consider model selection in a league
framework (model vs. model for explaining a certain observation). This is shown in Fig. \ref{fig:leagues}, where each square
indicates the value of the logarithmic evidence ratio obtained from the competition of
pairs of models. Red colors are associated to a model on the horizontal axis that
is preferable to a model in the vertical axis. Obviously, only half of the squares contain
relevant information.

Given that calculating a reliable estimation of the evidence is computationally very demanding, 
it is of interest to compare it with simpler proxies used for model comparison. The property
of such proxies is that they can be calculated very fast and it is not necessary to 
perform the multidimensional integral of the evidence. The Bayesian Information Criterion 
\citep[BIC;][]{schwarz_bic78} is one of the routinely used proxies. It is based on the crude approximation 
of gaussianity of the posterior with respect to the model parameters but it is extremely simple to calculate:
\begin{eqnarray}
\mathrm{BIC} &=& \chi^2_\mathrm{min} + k \ln N \\
\end{eqnarray}
where $k$ is the number of free parameters of the model, $N$ is the number of observed wavelength points
and $\chi^2_\mathrm{min}$ is:
\begin{equation}
\chi^2_\mathrm{min} = \sum_{j=1}^M \left( \frac{y_j(\widehat{{\mbox{\boldmath$\theta$}}})-d_j}{\sigma_j} \right)^2,
\end{equation}
the minimum value of the $\chi^2$ merit function attained for the vector of model parameters $\widehat{\mbox{\boldmath$\theta$}}$.
In the previous formula, $d_j$ is each one of the $M$ observed wavelength points (including the four Stokes
parameters), $\sigma_j$ is the standard deviation of the noise and $y_j(\widehat{{\mbox{\boldmath$\theta$}}})$ is
the Stokes profiles predicted for the vector of model parameters $\widehat{\mbox{\boldmath$\theta$}}$.

When comparing several models, the model with the smallest value of the BIC is the preferred one. 
We have verified that more than 80\% of the time
the BIC picks up the same model selected by the evidence ratio. Consequently, we suggest 
anyone carrying out inversions to compute the
value of the BIC for the selected model. This will facilitate model comparison in the
future.

\acknowledgements Financial support by the 
Spanish Ministry of Science and Innovation through projects AYA2010-18029 (Solar Magnetism and Astrophysical 
Spectropolarimetry) and Consolider-Ingenio 2010 CSD2009-00038 is gratefully acknowledged.

\bibliographystyle{asp2010}
\bibliography{/scratch/Dropbox/biblio}

\end{document}